# Path-Integral Monte Carlo And The Squeezed Trapped Bose-Einstein Gas

Juan Pablo Fernández[1,2] and William J. Mullin[1]

[1]*Department of Physics, University of Massachusetts, Amherst, MA 01003, U.S.A.*
[2]*Lebanon College, Lebanon, NH 03766, U.S.A.*

**Abstract.** Bose-Einstein condensation has been experimentally found to take place in finite trapped systems when one of the confining frequencies is increased until the gas becomes effectively two-dimensional (2D). We confirm the plausibility of this result by performing path-integral Monte Carlo (PIMC) simulations of trapped Bose gases of increasing anisotropy and comparing them to the predictions of finite-temperature many-body theory. PIMC simulations provide an essentially exact description of these systems; they yield the density profile directly and provide two different estimates for the condensate fraction. For the ideal gas, we find that the PIMC column density of the squeezed gas corresponds quite accurately to that of the exact analytic solution and, moreover, is well mimicked by the density of a 2D gas at the same temperature; the two estimates for the condensate fraction bracket the exact result. For the interacting case, we find 2D Hartree-Fock solutions whose density profiles coincide quite well with the PIMC column densities and whose predictions for the condensate fraction are again bracketed by the PIMC estimates.

**Keywords:** Bose-Einstein condensation; Two-dimensional Bose systems; Path-integral Monte Carlo simulations.

**PACS:** 03.75.Hh, 05.30.Jp, 05.70.Fh, 32.80.Pj

Though it has been rigorously proved that two-dimensional (2D) Bose-Einstein condensation cannot occur in the thermodynamic limit, the fact that it has been experimentally detected [1] in finite trapped gases is not too surprising. Nothing in principle forbids taking an anisotropic "pancake"-shaped trap and increasing the largest of its confining frequencies ($\omega_z \equiv \lambda \omega$, say, where $\omega \equiv \omega_x = \omega_y$) to an arbitrary degree, and it would be reasonable to expect some sort of quasi-2D behavior beyond a certain threshold. An ideal gas of $N$ atoms, for example, condenses at a higher temperature than its 3D counterpart ($T_{2D} = 0.829 N^{1/2} T_{3D}$), and thus it should be possible to condense a system by compressing it. A zero-temperature model based on the Thomas-Fermi approximation [1] predicts that the crossover for an interacting gas occurs at a compression ratio $\lambda = (225 m\omega\, a^2 N^2 / 32\hbar)^{1/3}$, where $m$ is the mass of each atom and $a$ is its $s$-wave scattering length. Beyond this point, the system acquires an effective two-dimensional coupling constant given by $g_{2D} = (\lambda m\omega / 2\pi\hbar)^{1/2} g_{3D}$, where $g_{3D} = 4\pi\hbar^2 a / m$; this result provides a characterization of a 2D system in terms of experimentally known parameters.

The smooth lines in Figure 1 show both the exact number density—integrated over $z$, expressed as a function of $\xi = (m\omega/\hbar)^{1/2} (x^2 + y^2)^{1/2}$, and multiplied by $2\pi\xi$ so that $\int d\xi\, N(\xi) = N$—of an ideal gas of 100 atoms at $T = 1.3\, T_{3D}$ compressed to $\lambda = 50$ and the density profile of a 2D gas at the same actual temperature, which here is $T = 0.728 T_{2D}$. The close agreement makes the curves almost indistinguishable. Both densities have been resolved into their condensate and noncondensate components.

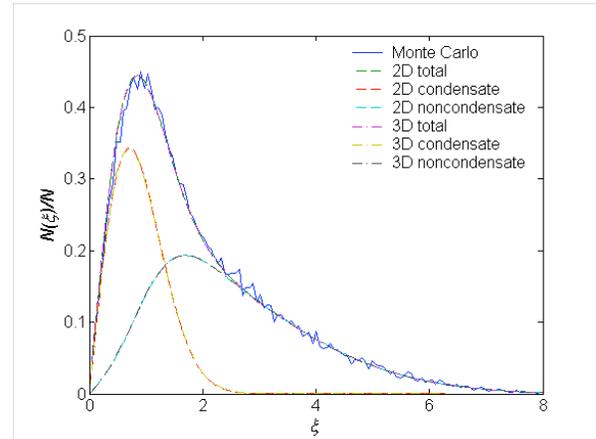

**FIGURE 1.** Surface number density of a compressed 3D ideal Bose gas and density of an identical 2D gas at the same temperature.

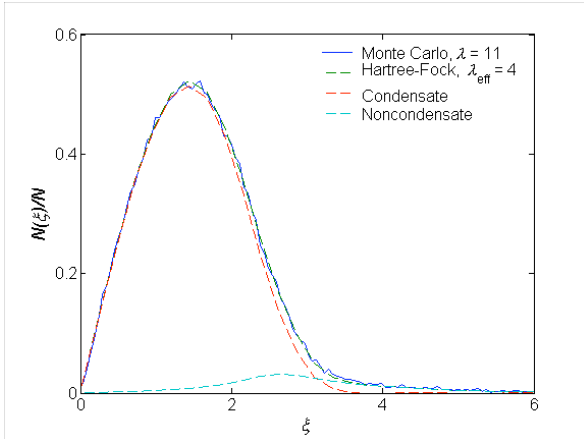

**FIGURE 2.** Monte Carlo number surface density and best-fit 2D profile of an interacting 3D Bose gas in a highly anisotropic trap.

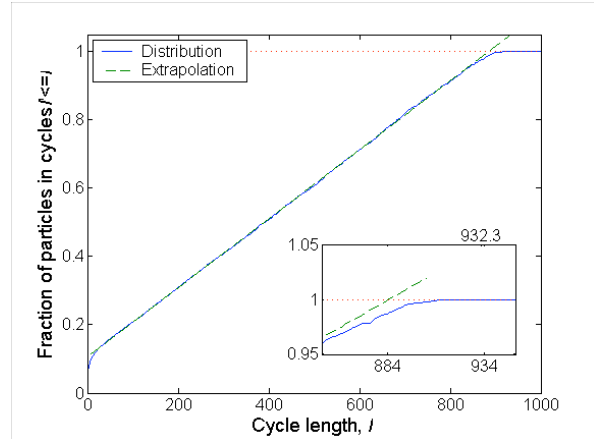

**FIGURE 3.** Fraction of atoms in cycles of lengths $l' \leq l$ for the gas in Fig. 2. To estimate $N_0$ we can take the first $l$ for which this is unity or extrapolate the linear segment up to 1.

The jagged line, which also agrees quite well with the total density profiles, is the result of a path-integral Monte Carlo simulation of the compressed system. The method, pioneered in this context by W. Krauth [2], calculates ensemble averages using as a weighting function the $N$-body density matrix of the system—expanded via Feynman path integrals into slices with tractable high-temperature actions and fully symmetrized to account for Bose statistics. The resulting high-dimensional integral and sum over $N!$ possible label reshufflings is sampled by generating multiparticle, multislice moves appropriate for the ideal gas and sifting them through the Metropolis algorithm. The simulation incorporates interactions between the atoms by solving exactly the quantum two-body problem for hard spheres of radius $a$ and using a pair-product approximation. The program outputs histograms of particles' positions, from which the number density can be inferred directly. Two ways of extracting the condensate fraction $N_0$ use the fact that every single configuration of the system may be characterized by a permutation that can in turn be decomposed into $c_1$ 1-cycles, $c_2$ 2-cycles, etc., which information the program can easily store and output; the presence of nonzero $c_l$'s for high values of $l$ signals the occurrence of BEC. Krauth's original method [2] reasons that the longest cycle must contain condensed atoms exclusively; a second method [3] assumes that, while the non-condensed particles occur only in small cycles, the condensate is composed in equal parts of atoms from all cycles up to a maximum size, accessible through extrapolation, that corresponds to one large permutation cycle involving the entire condensate. For example, in the ideal-gas simulation described above, the program yields $N_0 = 62$ and $N_0 = 39$ respectively, which bracket the exact value, $N_0 = 40$. (The uncertainty is large due to the small $N$ but decreases in larger systems.)

Figure 2 depicts the Monte Carlo number surface density of an *interacting* three-dimensional Bose gas of $N = 1000$ rubidium atoms (which should become 2D at $\lambda \geq 5.07$) at $T = 0.5316\,T_{3D}$, with a compression ratio $\lambda = 11$. The Monte Carlo result is superimposed on a pure-2D mean-field profile (broken again into condensed and thermal components) obtained by solving the finite-temperature Gross-Pitaevskii equation for the condensate with a minimization-collocation method and treating the thermal cloud in the Hartree-Fock approximation [4]. The only adjustable parameter is the effective anisotropy parameter in the coupling constant; the profile that best fits the Monte Carlo density in this case has $\lambda_{\text{eff}} = 4$. In general we find this effective compression factor to be lower than the Thomas-Fermi prediction; this is presumably a finite-temperature effect.

Figure 3 shows the two methods used to find the condensate fraction of the same system. The inset zooms in on the region of interest and displays Krauth's and our estimates (934 and 884, respectively) along with the fraction predicted by the best-fit 2D Hartree-Fock calculation (932.3). The Monte Carlo values again bracket the mean-field result. We conclude that there is a phenomenon resembling a condensation into a single state in a 3D system so compressed as to exhibit quasi-2D behavior.

## REFERENCES

1. A. Görlitz *et al.*, *Phys. Rev. Lett.* **87**, 130402 (2001).
2. W. Krauth, *Phys. Rev. Lett.* **77**, 3695 (1996).
3. W. J. Mullin, S. D. Heinrichs, and J. P. Fernández, *Physica B* **284–288**, 9 (2000).
4. J. P. Fernández and W. J. Mullin, *J. Low Temp. Phys.* **128**, 233 (2002); **138**, 687 (2005).